\documentclass[referee]{aa}

\usepackage{graphicx}
\usepackage{natbib}
\usepackage{latexsym}
\usepackage{amsmath}
\usepackage{amssymb}
\usepackage{amstext}
\usepackage{color}
\usepackage{psfrag}

\def\k{km s$^{-1}$}
\def\ks{km s$^{-1}$~}
\def\d{$^\circ$}
\def\m{$^\prime$}
\def\s{$^{\prime\prime}$}
\def\hh{$^{\mathrm h}$}
\def\mm{$^{\mathrm m}$}
\def\ss{$^{\mathrm s}$}
\def\cm3{cm$^{-3}$}

\def\2{$^{12}$CO}
\def\3{$^{13}$CO}
\def\msol{M$_\odot$}

\begin{document}

\title{Discovery of a dense molecular cloud towards a young massive embedded star in 30 Doradus\thanks{The observations
were made with the Swedish-ESO Submillimetre Telescope, SEST, which was operated jointly by the Swedish
National Facility for Radioastronomy and ESO.}
}

\author {M. Rubio \inst{1}
\and S. Paron \inst{2}
\and G. Dubner  \inst{2}
}
                                                                                                               
\institute{Departamento de Astronom\'\i a, Universidad de Chile, Casilla 36-D, 
	Santiago, Chile\\
	\email{mrubio@das.uchile.cl}
\and Instituto de Astronom\'{\i}a y F\'{\i}sica del Espacio (IAFE),
             CC 67, Suc. 28, 1428 Buenos Aires, Argentina}

\offprints{M. Rubio}

   \date{Received <date>; Accepted <date>}

\abstract{ The 30 Doradus region in the Large Magellanic Cloud is one of the most outstanding star 
forming regions of the Local Group and a primary target to study star formation in an environment 
of low metallicity. }
{In order to obtain a more complete picture of the not yet consumed or dispersed cool gas, we searched 
for line emission from molecular clouds that could be associated with molecular hydrogen emission detected 
in the region.}
{We obtained a high sensitivity \2 J=2--1 map with the 15-m SEST telescope, complemented by pointed
observations of \3 J=2--1 and CS J=2--1.}
{We report the discovery of a dense molecular cloud towards an embedded
young massive star at $\sim$ 20\s~($\sim$ 5 pc, at the distance of 50 kpc) northwest of R136, 
the compact massive central stellar cluster powering 
30 Doradus in the LMC, that could be triggering star formation in the surrounding molecular clouds. 
We derived a molecular mass of $\lesssim 10^4$ \msol, a linear radius of 3 pc, as an upper limit,  
and a mean density of $\gtrsim 10^{3}$ cm$^{-3}$ for
the cloud. The detection of CS J=2--1 emission line indicates larger densities, $\sim 10^{6}$ cm$^{-3}$.
The dense molecular cloud is associated with molecular 2.12 $\mu$m H$_{2}$ emission. We suggest that the observed 
molecular gas could be the remains of dense molecular material surviving the action of strong UV fields and winds 
in which the young massive star has formed.}{}

\titlerunning{A molecular clump and a young massive star in 30 Doradus}
\authorrunning{M. Rubio et al.}

\keywords {galaxies: Magellanic Clouds --- ISM: HII Region --- ISM: individual (30 Doradus) --- ISM: molecules}

\maketitle 

\section{Introduction}

30 Doradus in the Large Magellanic Cloud (LMC) is the brightest giant star forming
region among all galaxies in the Local Group.
It is 10 times more luminous than NGC3603, considered to be the most spectacular HII
region in our Galaxy powered by a central compact massive cluster containing over 2000 \msol. 

30 Doradus is powered by a dense central compact cluster, R136. This cluster 
was initially thought to be a supermassive star (1000 \msol) to explain 
the measured UV photons required to ionize the HII region. Several very massive
stars in R136 and other hot stars around \citep{walborn97} emit a very large flux 
of Lyman continuum photons which ionize most of the surrounding gas,
forming a super--giant HII region. The stellar winds and possibly
supernova explosions have eroded an asymmetric cavity, bounded
on one side (the NW half) by denser material, seen in particular as
molecular clouds \citep{joha98}.
Today we know that 
the cluster R136 consists of a compact dense core (0.1 pc in diameter), where more 
than 65 O3-type stars, the most luminous stars known, and more than 10 WR stars have been 
disentangled with {\it HST} observations \citep{massey98}.
R136 has a stellar mass of $10^5$ \msol~and a total luminosity of $\sim$ $10^8$ L$_{\rm \odot}$. 
The R136 cluster, the prototypical Super Star Cluster (SSC), 
is visible by naked eye at a distance of 50 kpc.  
 
In this sense, 30 Doradus is exceptional and the 
study of star formation and the molecular ambient in this HII region  
is crucial to understand the star forming processes around massive 
compact SSCs interacting with low metallicity molecular clouds, which are 
exposed to high UV radiation fields and strong winds from WR and luminous O stars. 
The physical conditions in this ISM resemble those that existed in the
early universe and thus can shed important light on the primeval
process of star formation.

30 Doradus has been extensively studied in optical and near infrared (NIR) wavelengths \citep{pogli95,rubio98}.
Several bright embedded infrared sources \citep{rubio98}, an H$_{2}$O maser and numerous nebular 
micro-structures as revealed by {\it Hubble Space Telescope (HST/WFPC2)} images  evidence
an active as well as recent and ongoing star formation (\citealt{walborn99} and references therein). 

\begin{figure*}[tt!]
\centering
\includegraphics[width=14cm]{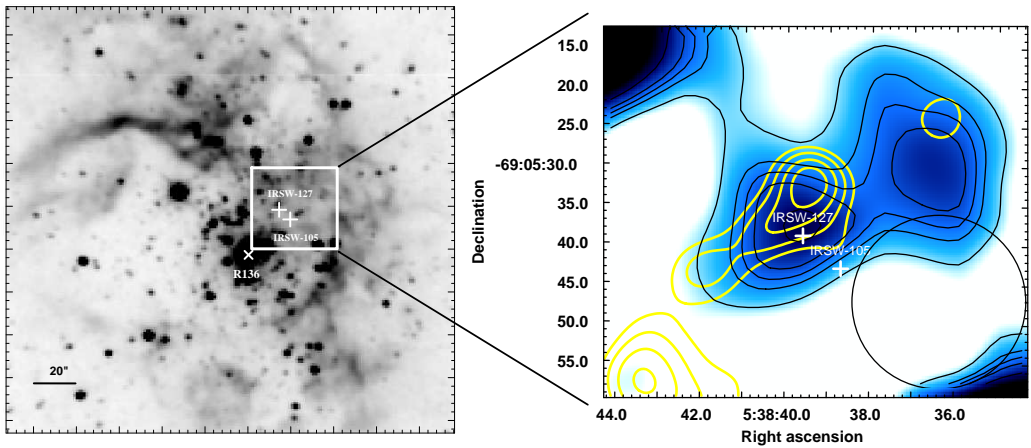}
\caption{{\it Left}: NIR  2.14 $\mu$m continuum image of 30 Doradus showing the position of IRSW-105,
IRSW-107 (crosses) and the center of R136 (the X). The box shows the extent of the map displayed
in the right panel.
{\it Right}: \2 J=2--1 emission integrated between 232 and 254 \ks towards the mentioned
IR sources. The black contour levels are 5, 6, 6.5, 7 and 8 K \k. The yellow contours
correspond to the H$_{2}$ near infrared emission. The black circle is the \2 J=2--1 24\s~beam.}
\label{clump}
\end{figure*}

In particular, \citealt*{rubio92} (RRG)
reported for the first time an isolated IR source (source W9)
located $\sim$ 20\s~northwest of R136. Based on the source position in the {\it J-H} vs {\it H-K} color-color 
diagram and its spectral energy distribution, the authors proposed that W9 was
most probably a young massive star.
Ground based infrared observations with high sensitivity and improved spatial resolution
allowed for the discovery of several new bright embedded IR sources associated with molecular clouds
in this region \citep{rubio98}.
All of the RRG sources were identified and 
some of them were resolved into 
several components. RRG-W9 was resolved into two sources, IRSW-127, a very bright object 
with the largest IR excesses among those in 30 Doradus ({\it K$_{S}$} $=$ 13.48, {\it K$_{S}$-H} $=$ 2.85),
and IRSW-105, displaced by $\sim$ 6\s.
These two IR sources  
show spatial coincidence (within $\sim$ 10\s) with the O-type stars P621 and P600
classified as O3--O6 V by \citet{walborn97}, who   
remarked that they might be very young objects on or near the zero-age main sequence.
As \citet{rubio98} suggest, the IR sources and the mentioned O-type stars likely represent
the same star-formation envents.

Star formation occurs in molecular clouds. These clouds are expected to be destroyed
by the violent interaction of the newly born massive stars.
The two IR sources were found at only 0.5 pc from the energetic massive cluster 
R136 and their proximity motivated us to study the molecular gas near the
massive and luminous stellar cluster R136.

In this work, we present \2 J=2--1, \3 J=2--1, CS J=2--1 and \2 J=1--0 
data towards IRSW-105 and IRSW-127.

\section{Observations}
 
High sensitivity observations in \2 J=2--1 (rest frequency: 230.538 GHz) 
were made between January and July, 1998 using the 15m Swedish-ESO Submillimetre 
Telescope (SEST). The observations were done in position switch mode with a fixed
reference off position free of CO emission. A narrow band AOS high-resolution (HRS) spectrometer with 
2000 channels, 80 MHz bandwidth and 41.7 kHz channel separation (corresponding to 0.054 \ks for the \2 J=2--1 line)
was used as back end.
At 230 GHz the beamwidth and main bean efficiency of the telescope were 24\s~and 0.60, respectively.
Calibration was done using the standard chopper wheel technique. The pointing accuracy, checked
during the observations on RDor, was better than 2\s.
Typical system temperatures were $\sim$ 200 K. The observations have an rms noise
of 0.07 K, achieved after 4 minutes of integration 
in each position, which notably improve the sensitivity reached by \citet{joha98} and \citet{ott08} in their 
\2 J=1--0 observations. Mapping was done with 10\s~spacings to produce fully
sampled maps. The data were reduced using CLASS\footnote{CLASS is a GILDAS software for reduction 
and analysis of (sub)-millimeter spectroscopic data. 
GILDAS, the ``Grenoble Image and Line Data Analysis Software'', 
is a collection of software developed by the Observatoire de Grenoble and IRAM. 
} 
and linear and in few cases, third order polynomia, were used for baseline fitting.
The spectra were smoothed to a velocity resolution of 0.25 \k.

The observations covered the region between the two strongest CO molecular clouds mapped in 30 Doradus 
(Cloud 10 and Cloud 6, \citealt{joha98})
where no CO emission was detected by \citet{joha98}. However, \citet{rubio98} observed concentrations of 
vibrational excited H$_{2}$ at 2.12 $\mu$m in this region, clearly calling for more sensitive CO observations. 
In this study, we restrict ourselves to the observations covering the
area near the central cluster R136. This region is shown in Figure \ref{clump} (left) with 
a 2.14 $\mu$m continuum image of the 30 Doradus Nebula. Two crosses show the position of
the IR sources and an X indicates the center position of R136. The box indicates the region 
that is analysed in the \2 J=2--1 line. 
 
We also obtained \3 and CS J=2--1 spectra towards the position of the strongest \2 J=2--1 emission
in the area, RA $=$ 5\hh 38\mm 39.5\ss, dec $=-$69\d 05\m 37.5\s~(J2000). The telescope beam at the 
CS J=2--1 transition is 45\s~with a beam efficiency of 0.70. These spectra were 
reduced in a similar way as the \2 J=2--1 observations. 
A \2 J=1--0 spectrum towards this position was obtained from \citet{joha98}.

\section{Results and Discussion}

\begin{figure*}[tt!]
\centering
\includegraphics[height=12cm]{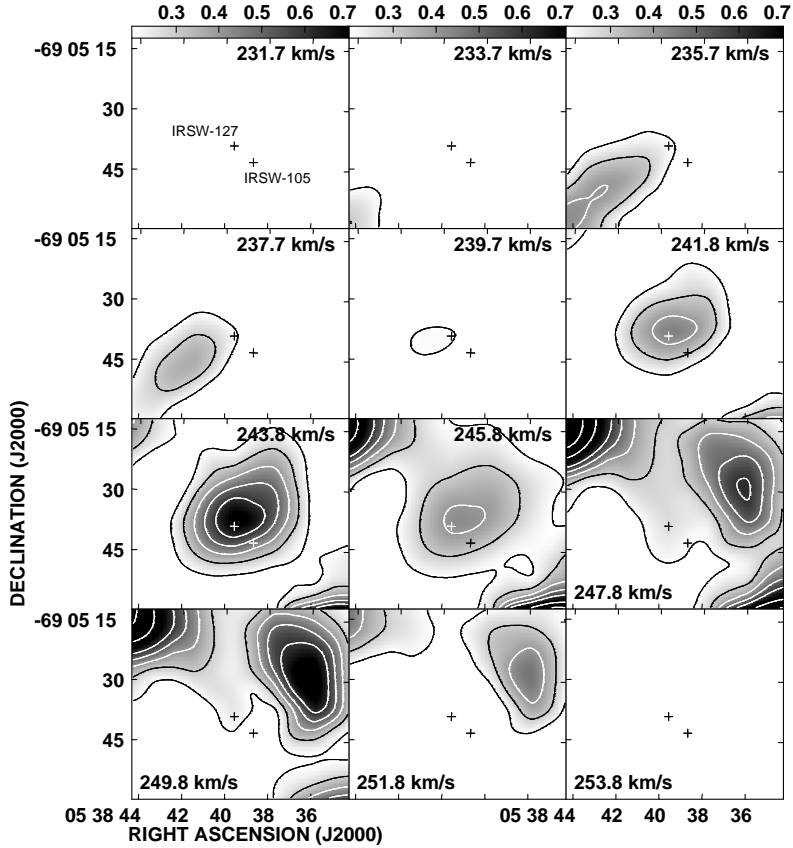}
\caption{\2 J=2--1 emission integrated over velocity intervals of $\sim$ 2 \k. The contour levels 
are 0.2, 0.3, 0.4, 0.5, 0.6, 0.8 and 1 K \k. The IR sources IRSW-105 and IRSW-127 are indicated by crosses.}
\label{panel}
\end{figure*}

Figure \ref{clump} (right) displays the velocity integrated \2 J=2--1 emission
between 232 and 254 \ks covering an area of approximately 50\s~$\times$ 50\s~around
IRSW-127 (W9) and IRSW-105, which are indicated by crosses.
The \2 J=2--1 map likely shows two CO clumps. The strongest one has its maximum very near IRSW-127 
and on its SW border lies IRSW-105. The second CO clump is seen towards the northwest and does not show 
any IR sources.

Previous studies of the ISM in 30 Doradus have shown that H$_{2}$ structures appear intimately associated with 
bright IR sources, embedded O stars, and dense molecular clumps exposed to strong stellar radiation
field \citep{pogli95,rubio98}. In Figure \ref{clump} (right), we have included in yellow contours the 2.12 $\mu$m 
molecular hydrogen emission extracted from \citet{rubio98}. 
The H$_{2}$ emission and the CO contours show a similar spatial distribution, with 
the IR source IRSW-127 associated with the brightest H$_{2}$ and CO emission. 

Table 1 presents the molecular parameters of the strongest CO clump displayed in Figure \ref{clump} (right), 
which can also be appreciated in the $\sim$ 244 \ks velocity channel
map in Figure \ref{panel}. 
T$_{mb}$ is the peak brightness temperature, V$_{LSR}$ the central velocity, 
$\Delta v$ the line width (FWHM) and $I$ the integrated line intensity. 
Errors are formal 1$\sigma$ value for the model of the Gaussian line shape. 

To determine the size of the molecular clump we applied $ R = \left( \frac{N}{\pi}\right)^{1/2} \Delta s$, where 
$\Delta s$ is the linear grid spacing and $N$ is the number of positions 
at which the cloud is detected above $\sim$5~$\sigma_{\rm rms}$. We obtain a radius of $\sim$ 3 pc. 
Since the angular size of the considered structure is comparable to the angular resolution of the present
observations, a linear radius of 3 pc can be derived as an upper limit for the size of this clump.
We derived the CO luminosity  L$_{{\rm CO}} = 120 $ K \ks pc$^{2}$, using
L$_{{\rm CO}} =  I_{\rm CO}$ $ N (\Delta s)^{2}$, 
where $I_{\rm CO} = \int{T_{mb}~dv}$.

\begin{table}[h]
\caption{\2 J=2--1 molecular parameters of the strongest CO clump displayed in Figure \ref{clump} (right). 
T$_{mb}$ is in K, V$_{c}$ and $\Delta v$ are in \k, 
$I_{\rm CO}$ in K \k, R in pc and L$_{\rm CO}$ in K \ks pc$^{2}$. }
\label{cloudparam}
\centering
\begin{tabular}{cccccc}
\hline\hline
 T$_{mb}$ & V$_{c}$ & $\Delta v$  &   $I_{\rm CO}$  & R & L$_{\rm CO}$ \\
\hline
0.65 $\pm0.05$ & 244.40 $\pm0.40$ & 4.40 $\pm0.50$ & 4.00 $\pm0.20$ & 3 & 120 \\
\hline
\end{tabular}
\end{table}

In Figure \ref{panel} we present channel maps of the velocity integrated \2 J=2--1 emission in steps 
of $\sim$ 2 \ks towards IRSW-127 and IRSW-105.
The two IR sources are located at the position of 
the peak and the SW border of \2 J=2--1 emission in the $\sim$ 244 \ks velocity channel 
map. The spatial coincidence of IRSW-127 with the molecular clump peak and the large
IR excess of this source, strongly suggests that IRSW-127 is embedded in this molecular clump. 
CO emission is also seen at lower and higher velocity components in a different spatial location as that
of the $\sim$ 244 \ks component.  
At lower velocities, the CO emission corresponds to an elongated 
structure extending from southeast to the center of the region. At higher velocities, the CO emission is
associated with other molecular structure that contains the northwestern clump 
and others clumps farther away. This is evidence for a high degree of clumping in the region.

\subsection{Physical properties}

\begin{figure}[h]
\centering
\includegraphics[height=15cm]{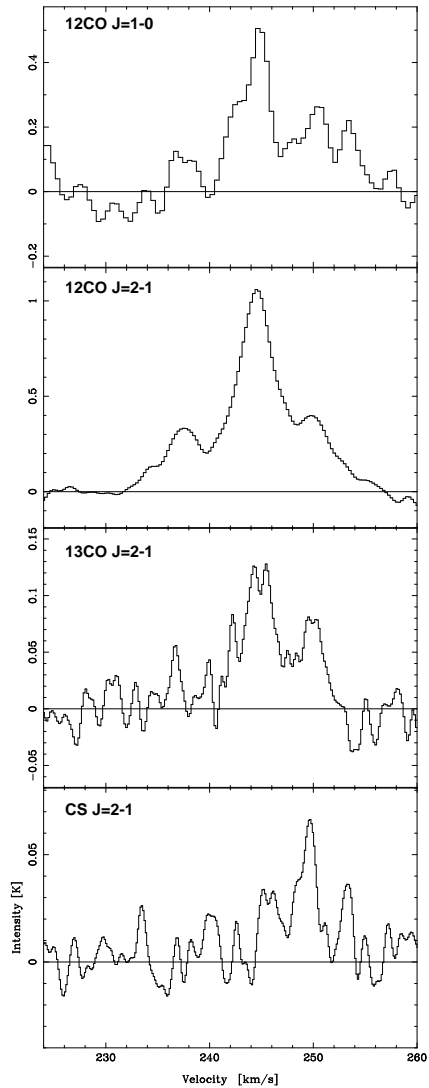}
\caption{ \2 (1--0), \2 (2--1), \3 (2--1) and CS (2--1) spectra observed at RA $=$ 5\hh 38\mm 39.5\ss,
dec $=-$69\d 05\m 37.5\s~(J2000). The intensity scale which is in T$_{mb}$ has been adjusted to better show the emission
line. }
\label{espectros}
\end{figure}

Figure \ref{espectros}  
shows the spectra of the \2 J=1--0 and J=2--1, \3 and CS J=2--1 transitions obtained towards 
the  position of the strongest \2 J=2--1 emission; RA $=$ 5\hh 38\mm 39.5\ss, dec $=-$69\d 05\m 37.5\s~(J2000). 
The spectra were hanning smoothed and are presented as histograms for a better display. 
In all the CO transitions three velocity components are seen, at 237 \k, 244 \k, and 249 \k, respectively,
while the CS J=2--1 spectrum shows emission only at the two higher velocity components. Table \ref{param} summarizes
the line emission parameters as obtained from gaussian fit to each spectra.
 
\begin{table}[h]
\tiny
\caption{Observed parameters from the spectra at the peak position RA $=$ 5\hh 38\mm 39.5\ss,
dec $=-$69\d 05\m 37.5\s~(J2000).}
\label{param} 
\centering
\begin{tabular}{ccccc} 
\hline\hline 
Line emission & V$_{c}$ & T$_{mb}$ & $\Delta v$  & $I$\\ 
         & (\k)   & (K)     &  (\k)      & (K \k)           \\ 
\hline
\2 J=1--0 & 237.6 $\pm$0.9 & 0.11 $\pm$0.04  & 2.1 $\pm$0.2 & 0.30 $\pm$0.12 \\
          & 244.9 $\pm$0.8 & 0.50 $\pm$0.06  & 3.8 $\pm$0.4 & 1.82 $\pm$0.25 \\
          & 251.4 $\pm$0.9 & 0.25 $\pm$0.05  & 5.6 $\pm$0.2 & 1.33 $\pm$0.25 \\
\hline 
\2 J=2--1 & 237.8 $\pm$0.9 & 0.30 $\pm$0.05  & 5.7 $\pm$0.5 & 1.40 $\pm$0.25 \\
          & 244.4 $\pm$0.7 & 1.00 $\pm$0.08  & 4.0 $\pm$0.5 & 5.20 $\pm$0.25 \\
          & 249.7 $\pm$0.8 & 0.37 $\pm$0.05  & 5.5 $\pm$0.7 & 1.60 $\pm$0.22 \\
\hline
\3 J=2--1 & 236.7 $\pm$0.3 & 0.08 $\pm$0.02  & 0.4 $\pm$0.3 & 0.07 $\pm$0.02 \\
          & 244.4 $\pm$0.3 & 0.13 $\pm$0.01  & 3.1 $\pm$0.6 & 0.40 $\pm$0.04 \\
          & 249.8 $\pm$0.2 & 0.07 $\pm$0.01  & 2.1 $\pm$0.6 & 0.16 $\pm$0.03 \\
\hline
CS J=2--1 & 245.7 $\pm$0.4 & 0.03 $\pm$0.01 & 1.7 $\pm$0.6 & 0.07 $\pm$0.02 \\
          & 249.5 $\pm$0.2 & 0.07 $\pm$0.01 & 2.1 $\pm$0.6 & 0.12 $\pm$0.03 \\
\hline
\end{tabular}
\end{table}

We determined the line ratios of the strongest CO clump at the velocity component $\sim 244$ \k.
These ratios are presented in Table \ref{ratios}, 
which includes the intensity ratio and the velocity integrated line intensity ratio at the
peak position convolved to the same beam size. 

On the other hand, integrating the \2 J=1--0 data from \citet{joha98} around $\sim 244$ \k, 
we found a weak emission at the position of our CO clump. 
Convolving the \2 J=2--1 map to the J=1--0 angular resolution we obtained 
the \2 J=2--1 and J=1--0 integrated line ratio ($R_{2-1/1-0}$). 
For the CO clump we obtained $R_{2-1/1-0} \sim 1.5$.

\begin{table}[h]
\tiny
\caption{Line ratios for the $244$ \ks component at the peak position. The \2 J=2--1 emission was convolved to the 
\2 J=1--0 and CS J=2--1 beams, respectively.}
\label{ratios}
\centering
\begin{tabular}{cccc}
\hline\hline
Peak Line Ratio & $\frac{^{12}\rm{CO}(2-1)}{^{12}\rm{CO}(1-0)}$ & $\frac{^{12}\rm{CO}(2-1)}{^{13}\rm{CO}(2-1)}$ & $\frac{^{12}\rm{CO}(2-1)}{\rm{CS}(2-1)}$ \\ 
\hline
Intensity: T$_{mb}$    & 1 $\pm0.2$  & 8 $\pm1.2$  & 15 $\pm7$ \\ 
Vel. Integ.: $\int{T_{mb}~dv}$   & 1.4 $\pm0.3$  & 13 $\pm2$  & 38 $\pm30$ \\
\hline
\end{tabular}
\end{table}

The value obtained in the isotopic 
line integrated ratio is in agreement with those for molecular gas in 
other star forming regions in the Magellanic Clouds \citep{israel03X},
and compatible with a low metallicity and strong radiation field. 
The multiple peak structure and the presence of wings in the spectra may indicate that 
the main contribution to the gas motion could be a turbulent 
velocity field \citep{falgarone94,garay02}.

The detection of CS J=2--1 towards this molecular clump implies the presence of dense gas.
Generally, this `high-density' tracer indicates densities $\sim$ 10$^{5}$ cm$^{-3}$ with 
an upper limit of a few $\times 10^{6}$ cm$^{-3}$ \citep{heik99,wang09}.
The CS J=2--1 line traces the dense interior of the clump, while the CO lines trace the external and 
less dense envelopes. The angular resolution of the CS J=2--1 observations is a factor of two 
lower than the \2 J=2--1 observation, which implies that the CS spectra covers the total 50\s~$\times$ 50\s~studied area. 
Thus the CS emission could be coming from several dense regions. 
The most intense CS component is seen at $\sim 249$ \ks and a second
weaker component is at $\sim 245$ \k, very close to $\sim 244$ \ks where the CO main component is seen.  
These different velocity components seen in the CS spectra could be an another evidence 
for clumpiness of the region.

We determine the optical depth of the CO clump considering an 
isotopic abundance ratio ([\2/\3]) between 40 and 50. 
\citet{heik99} obtained [\2/\3] ratios of 38 and [HCO$^{+}$/H$^{13}$CO$^{+}$] ratios of 
$35\pm21$ in 30Dor-10, and in the SMC N27 cloud, they obtained isotopic abundance ratios of 35 and 40-90, respectively. 
Recently, \citet{wang09} obtained the isotope ratio  
[$^{12}$C/$^{13}$C] $\sim 49\pm5$ in the star forming region N113 in the LMC. 
Assuming  T$_{ex} = 10$ K, that the surface filling factors 
and the excitation temperatures of the \2 and \3 emissions are equal, and using 
the ratio of the peak temperatures, we calculate the \2 and \3 J=2--1 optical depths for this cloud. We 
obtain $\tau^{12} \sim 6$ and $\tau^{13} \sim 0.13$, respectively.  
This result would not be significantly altered if T$_{ex}$ were as larger as 30 K.
The \2 J=2--1 line appears optically thick while the emission 
in the \3 J=2--1 line is optically thin. The ratios obtained from our CO clump
are similar to those obtained in N159 \citep{pineda08}. Different beam filling factors
play a minor role on the line ratios, even in a clumpy medium \citep{gierens92,pineda08}.  

\subsection{Molecular mass}
                          
We have used three independent ways to estimate the molecular mass of the 
discovered molecular clump. First, we assume that the molecular clump is in virial equilibrium.
Assuming that the clump has a spherical shape, a radius of 3 pc, which 
is an upper limit, 
we estimate M$_{\rm{vir}} \lesssim 1 \times 10^{4}$ \msol,
using the constant B $=$ 190 for 
a density profile of $\rho(r) \propto 1/r$ \citep{mac88}. 
Secondly, we estimated the molecular clump mass from the observed \2 J=2--1 luminosity
using an H$_{2}$ -- \2(1--0) conversion factor of $X \sim 7 \times 10^{20}$ cm$^{-2}$ K$^{-1}$ km$^{-1}$ s,  
2.5 times the Galactic value, as determined for this LMC region 
(\citealt{garay02,israel03X} and reference therein). 
We multiply this conversion factor by the integrated line ratio $R_{2-1/1-0} = 1.5$ 
obtained for the studied CO clump obtaining a molecular mass of $M_{{\rm CO}} \lesssim 2.7 \times 10^{3}$ \msol.  
Thirdly, we derived the mass using the \3 J=2--1 emission and by assuming LTE conditions.
We assume optically thick $^{12}$CO J=2--1 emission and a beam filling factor of 1, which may not
be completely true but allows us to make a first initial guess. 
We derived an optical depth, $\tau^{13} \sim  0.12$
and using the fractional abundance of [H$_{2}$/\3] $\sim 1.5 \times 10^7 $ 
for this region \citep{heik99} we obtain M$_{\rm{LTE}}$ $\lesssim$ 1.5 $\times 10^4$ \msol. 
As a result, we derive a number density of H$_{2}$
molecules of n$_{\rm{vir}} \gtrsim   2.0 \times 10^{3}$ cm$^{-3}$,  n$_{\rm{CO}} \gtrsim 0.5 \times 10^{3}$ cm$^{-3}$
and  n$_{\rm{LTE}} \gtrsim 2.8 \times 10^{3}$ cm$^{-3}$.

Taking into account that the mass values are upper limits, in what follows we perform a rough comparison between
them. The virial mass and the LTE mass are similar, being the LTE mass slightly larger, implying that the 
molecular clump may be approximately in virial equilibrium.
The mass obtained from the observed CO luminosity is approximately one 
order of magnitude lower than the virial and LTE masses. 
Such difference may be indicative of
the strong photo-dissociation that this molecular clump has suffered
due to its location in extreme environmental conditions. 
The molecular clump may have a large H$_{2}$ envelope where most of
the mass is found with only a dense CO core which has survived from photo-dissociation \citep{leq94}.
Another possibility to reconcile the masses estimated in different ways is by assuming  
a conversion factor 13 times
larger than the Galactic value, as obtained by \citet{garay93} for molecular clouds in the 30 Doradus halo.
Using this conversion factor we obtain a molecular mass from the observed CO luminosity of 
$M_{{\rm CO}} \lesssim 1.4 \times 10^{4}$ \msol, in agreement with the LTE mass.

\section{Conclusions}

We discovered a molecular clump located $\sim$ 20\s~northwest of the center of the compact cluster R136, 
where the IR source IRSW-127, a young massive star, is likely embedded. This source appears very strongly 
correlated with the molecular maximum. 
We derived a clump radius of 3 pc as an upper limit, a molecular mass 
of $\lesssim 10^4$ \msol, and a density of $\gtrsim 10^{3}$ cm$^{-3}$ from 
the \2 J=2--1 emission. The $R_{2-1/1-0} \sim 1.5$, higher than
the average value of $R_{2-1/1-0} \sim 0.8$ in a 9 arcmin beam, is consistent with optically thick, 
thermalized, dense gas  ($\geq 10^{3}$ cm$^{-3}$) ready to form stars as pointed out by \citet{sorai01}. 

The molecular clump lies in a region of H$_{2}$ emission.
\citet{pogli95} suggest that the clumpy H$_{2}$
emission that they observed in 30 Doradus may be coming from photo dissociation regions (PDRs) and   
could be produced in dense $\sim$ 10$^{6}$ cm$^{-3}$ molecular clumps exposed to the 
stellar radiation field. The molecular clump studied in this work is exposed to an UV field of $\sim$ 6000 times 
the average Galactic UV radiation 
field as determined using the Starburts99 code \citep{leit99} for a 2 Myr cluster with a metallicity of 0.008 
at the distance of 20\s~of R136 (Guzm\'an, private communication). Thus, the clump could have a large  
envelope of self-protected molecular gas.

Our detection of CS J=2--1 emission confirms the presence of dense gas, $\sim$ 10$^{6}$ cm$^{-3}$, in the molecular
clump. It is detected at $\sim 245$ \ks, very close to the velocity component where 
the \2 J=2--1 emission is stronger and spatially correlated with the two IR sources. 
Although the CS J=2--1 emission
is stronger at the $\sim 249$ \ks velocity component, the CO emission at this velocity does not show spatial coincidence
with the IR sources. Thus, this CS emission could come from another dense clump where massive star formation has not 
yet occurred, or if it has occurred, the possible IR sources would not been detected by existing surveys because they could
suffer an even larger extinction than that of IRSW-127. Higher angular resolution observations of this high density
tracer would be needed to study the dense gas structure of the region.

As mentioned above, the CS velocity structure shows evidence of clumpiness, which agrees with previous results 
for the 30 Doradus Nebula 
\citep{joha98,heik99} and is compatible with star forming activity. We find additional evidence
of this activity: (1) the LTE mass (M$_{\rm LTE}$) 
is slightly larger than the virial mass (M$_{\rm vir}$), which according to \citet{kawamura98} and \citet{tachi00} 
could be interpreted as an indication of star formation, and (2) the isotopic line integrated ratio (\2/\3) 
agrees with those found in different star forming regions in the Magellanic Clouds.

We conclude that the discovered molecular clump could be the remain of molecular material which has
survive the interaction of the strong winds and UV field of R136.
This CO clump is an example
of a dense molecular cloud with a large fraction of its mass in H$_{2}$ and not in CO as this molecule
has suffered strong photo-dissociation from the UV radiation flux of the massive central cluster R136.

We confirm the suggestions that possible energetic winds and strong UV radiation from the stellar 
cluster R136 could be triggering the star formation in the surrounding molecular clouds, and that 
IRSW-127 could be a newly formed young massive star embedded in a dense molecular clump.

\begin{acknowledgements}
M.R. is supported by the Chilean {\sl Center for Astrophysics}
FONDAP No. 15010003. 
S.P. and G.D. are member of the {\sl Carrera del 
investigador cient\'\i fico} of CONICET, Argentina. 
We would like to thank Viviana Guzman for the computation of the
UV radiation field from R136.
This work was partially supported by the CONICET 
grant 6433/05, UBACYT A023 and ANPCYT PICT 04-14018. M.R. and S.P. acknowledge support from FONDECYT N\d~1080335. 

\end{acknowledgements}

\bibliographystyle{aa}  
\bibliography{bib-30dor}
\IfFileExists{\jobname.bbl}{}
{\typeout{}
\typeout{****************************************************}
\typeout{****************************************************}
\typeout{** Please run "bibtex \jobname" to optain}
\typeout{** the bibliography and then re-run LaTeX}
\typeout{** twice to fix the references!}
\typeout{****************************************************}
\typeout{****************************************************}
\typeout{}
}

\end{document}